\documentclass[final,5p,times,twocolumn]{elsarticle}

\usepackage{graphicx}
\usepackage{bm}
\usepackage{lineno}
\usepackage{lineno, blindtext}
\usepackage{subcaption}
\usepackage{multirow}
\usepackage{amssymb}

\journal{Journal of Alloys and Compounds}

\begin{document}


\title{Electrical properties of SmB$_6$ thin films \\prepared by pulsed laser deposition 
from a stoichiometric SmB$_6$ target}

\author{Marianna Batkova\corref{cor1}}
\ead{batkova@saske.sk}

\author{Ivan Batko\corref{cor2}}
\address{Institute of Experimental Physics, Slovak Academy of Sciences, Watsonova 47, 040~01~Ko\v {s}ice, Slovakia}

\author{Feliks Stobiecki}
\author{Bogdan Szyma\'{n}ski}
\author{Piotr Ku\'{s}wik }
\address{Institute of Molecular Physics, Polish Academy of Sciences, ul. Smoluchowskiego 17, 60-179 Pozna\'{n}, Poland}

\author{Anna Mackov\'a}
\author{Petr Malinsk\'y}
\address{Nuclear Physics Institute of Academy of Sciences of the Czech Republic, 25068 Rez Near Prague, Czech Republic}
\address{Department of Physics, Faculty of Science, J.E. Purkinje University, Ceske Mladeze 8, 400 96 Usti nad Labem, Czech Republic} 


\cortext[cor1]{Corresponding author}

\begin{abstract}

Possible existence of topologically protected surface in samarium hexaboride has created a strong need 
	for investigations allowing to distinguish between properties coming from the surface states
	and those originating in the (remaining) bulk.
Studies of SmB$_6$ thin films represent a favorable approach allowing well defined variations
of the bulk volume that is not affected by surface states. 
Moreover, thin films are highly desirable for potential technology applications. 
However, the growth of SmB$_6$ thin films is accompanied by technology problems, 
which are typically associated with maintaining the correct stoichiometry of samarium and boron.
Here we present feasibility study of SmB$_6$ thin film synthesis 
by pulsed laser deposition (PLD) from a single stoichiometric SmB$_6$ target.
As proved by Rutherford Backscattering Spectrometry (RBS), we succeeded to obtain the same ratio 
of samarium and boron in the films as that in the target.
Thin films revealing characteristic electrical properties of (crystalline) SmB$_6$ 
			were successfully deposited on MgO, sapphire, and glass-ceramics substrates, 
			when the substrates were kept at temperature of 600~$^\circ$C  during the deposition. 
Performed electrical resistance studies 
				have revealed that bulk properties of the films are only slightly affected by the substrate.
Our results indicate that PLD is a suitable method 
			for complex and intensive research of SmB$_6$ and similar systems.
\end{abstract}

\begin{keyword}
\texttt{rare earth alloys and compounds\sep thin films\sep vapor deposition\sep electrical transport\sep valence fluctuations\sep Rutherford backscattering, RBS}
\end{keyword}


\maketitle

\biboptions{sort&compress}

\section{Introduction}

	 Heavy fermion semiconductor samarium hexaboride, SmB$_6$,
				attracts  attention of researchers for more than half a century. 
Its fascinating physical properties are very interesting not only because of 
			long-lasting fundamental questions,
				but also from an application point of view 	\cite{Wachter93, Riseborough00, Yong14, Yong15}. 
				Electrical resistivity of high-quality SmB$_6$  samples below 50~K shows a rapid increase with decreasing temperature,
				and surprisingly, it tends to saturate at very high residual value $\rho_0$ 
				at the lowest temperatures, below 4~K
				\cite{Allen79, Batko93, Cooley95}.
	Recently, SmB$_6$  has been predicted to be a prototype of topological Kondo insulator \cite{Dzero10, Dzero12, Lu13, Alexandrov13}, 
			and the strange low-temperature behavior is being explained by the metallic surface, 
			whereas nontrivial topological surface states \cite{Dzero10, Dzero12, Lu13, Alexandrov13}, 
			as well as ”trivial” polarity-driven ones \cite{Zhu13} were proposed to exist there.
	Valence-fluctuation induced hopping transport \cite{BaBa14} is another possibility 
			to explain the puzzling electrical transport in SmB$_6$ at the lowest temperatures \cite{BaBa14}. 
In principle, all the mentioned scenarios could even coexist, nevertheless, the highest attention 
			is paid to the SmB$_6$ being a possible topological Kondo insulator.   
Topological insulators represent an interesting new class of materials that 
		is promising for both fundamental and applied research \cite{Yong14}.
The surface states being robust against scattering from non-magnetic impurities 
			and displaying spin-momentum locking predict a possible effective use 
			of topological insulators in spintronics, magneto-electrics, 
			and quantum computing \cite{Yong14, Yong15, Qi11, Hasan10}. 
However, utilization of the surface transport 
				is often complicated by residual conductivity of the bulk  \cite{Yong14}.
 Advantage of SmB$_6$ should be that it is expected to be a \emph{true} topological insulator,
			i.e. with \emph{zero} residual bulk conductivity, which is highly 
			desired from application point of view \cite{Yong14}.
As reported recently, 
		the bulk electrical resistivity of SmB$_6$
			is thickness independent \cite{Petrushevsky17, Syers15}, and the electrical conduction 
			of SmB$_6$ samples with different thickness can be adequately described 
		by varying relative ratio of the bulk and surface contribution 
		to the total conduction \cite{Syers15}.
For further progress in both,
			understanding the underlying physics  and development of possible applications,
			ongoing research is needed and 
reproducible fabrication of well defined SmB$_6$ thin films allowing 
			to control 	relative contributions from the bulk and the surface
						is highly desired. 
	However, preparation of stoichiometric SmB$_6$ thin films is not straightforward.
	Recently, semi-epitaxial SmB$_6$ thin films were prepared by the molecular
				beam epitaxy with the obtained ratio	of B to Sm  between 4.9 and 5.7 \cite{Shishido15}. 
	Also two another groups reported a failure in attempts to deposit SmB$_6$ 
			thin films via pulsed laser deposition (PLD)  using a single SmB$_6$ target 
			\cite{Petrushevsky17, Yong14}.
	As reported there, one of the main difficulties at depositing SmB$_6$  
			is escape of boron  \cite{Petrushevsky17}, which leads to boron deficient films.
	To overcome this issue and achieve the correct stoichiometry, either co-sputtering 
			of SmB$_6$ and boron targets \cite{Yong14, Yong15} 
			(using the combinatorial composition-spread approach \cite{Jin13})
			or alternating ablation of SmB$_6$ and boron targets \cite{Petrushevsky17} 
			was finally used by those groups. 
On the other hand, it is worth to mention that almost 30 years ago SmB$_6$ thin films
			were prepared by evaporating the bulk material onto glass-ceramic substrates,
			while those films revealed electrical properties as typical 
			for semiconducting behavior of SmB$_6$, 
			as well as metalic-like conduction associated by the authors 
			with the surface properties \cite{Batko90}.
The mentioned work \cite{Batko90} indicates that it could be possible to obtain SmB$_6$
			thin films by deposition 	only from a single stoichiometric SmB$_6$ target.
Our aim has been to prove possibility to prepare SmB$_6$ thin films 
			from a single SmB$_6$ target by PLD technology, which is widespread in laboratory research 
			due to easy adaptation of deposition equipment and variability 
					of applied deposition parameters. 
					Moreover, among methods for thin film deposition the PLD is considered to achieve
					very close correspondence between the target composition and the deposited film.  
In this work  we report feasibility studies of SmB$_6$ thin film synthesis 
				by PLD from a single stoichiometric SmB$_6$  target 
				on several types of substrates and compare their properties with those of the target.

\section{Experimental} 
		SmB$_6$ thin films were deposited by PLD from a commercially available polycrystalline SmB$_6$ target
		(Testbourne, England). 
Nominal purity of the material was 99.5~$\%$, while the most contained contaminants reported were 
		Si (0.02~$\%$), Al (0.01~$\%$), and Fe (0.01~$\%$). 
The target was of disc geometry (25~mm in diameter, 5~mm thick). 
		Nd: YAG laser was used to generate laser pulses of wavelength 355~nm and 532~nm 
		at frequency of 2~Hz or 5~Hz. 
The 10~ns pulses with energy from the interval of 4 - 60~mJ were used to obtain the growth rate
			in the range of $0.1-0.5$~\AA/s. 
The base pressure in the deposition chamber was 5 $\times$ 10$^{-8}$~mbar.
Several types of substrate material were used: glass, MgO single crystal, sapphire, 
			and glass-ceramics (GC), namely, the non-polished ultra-low expansion GC 
			CLEARCERAM\textsuperscript{\textregistered}-Z (Ohara, Japan). 
Orientations of single crystal substrates were (001) and (012) for MgO and sapphire, respectively.
The substrate temperature during the deposition was defined by the temperature of substrate holder that was kept
				at ambient  temperature (glass, GC), or heated to 600~$^\circ$C (MgO, sapphire, GC).

Elemental composition of the thin films deposited on glass substrates (at ambient temperature) was determined by Rutherford Backscattering Spectrometry (RBS) 
				at Nuclear Physics 	Institute of the Czech Academy of Sciences on Tandetron~4130~MC accelerator. 
Ion beams of He$^{+}$ and H$^{+}$ ions with an energy of 2~MeV were used, 		
		while the ion beam 
		incoming angles were 0$^\circ$ and 45$^\circ$. 
			The glancing ion beam incoming angle was chosen in case of very thin layers characterization to increase the depth resolution.
Detection of backscattered ions was performed by the ion implanted detector ULTRA ORTEC
			in IBM geometry with scattering angle of 170$^\circ$. 
The detector was placed at the angle of 10$^\circ$ with respect to the incident beam. 
The RBS spectra were evaluated using the  SIMNRA software \cite{SIMNRA}.

Electrical resistance was measured by four-probe AC technique using standard 
		Van der Pauw geometry for samples deposited on substrates of square geometry (MgO, sapphire, GC), 
		or using the standard "in-line" geometry of contacts for thin films deposited on glass of oblong shape and the target.
AC resistance was measured using SR830 Lock-In Amplifier (Stanford Research Systems, USA)
		in combination with AMS220 Voltage Controlled Current Source (ANMESYS, Slovakia). 
Measurements were done in He$^4$ cryostat in temperature range of $4-300$~K.
		Temperature was determined utilizing calibrated Cernox~1050 and Pt-103 thermometers (Lake Shore Cryotronics, USA).

\section{Results and discussion}

In the first attempt, thin films were deposited on glass substrates (by PLD 
		from the single SmB$_6$ target), while the substrates were kept at ambient 
		temperature during the deposition.

\begin{figure}[h]
\centering
   \begin{subfigure}[h]{1\columnwidth}
   \includegraphics[width=0.95\columnwidth]{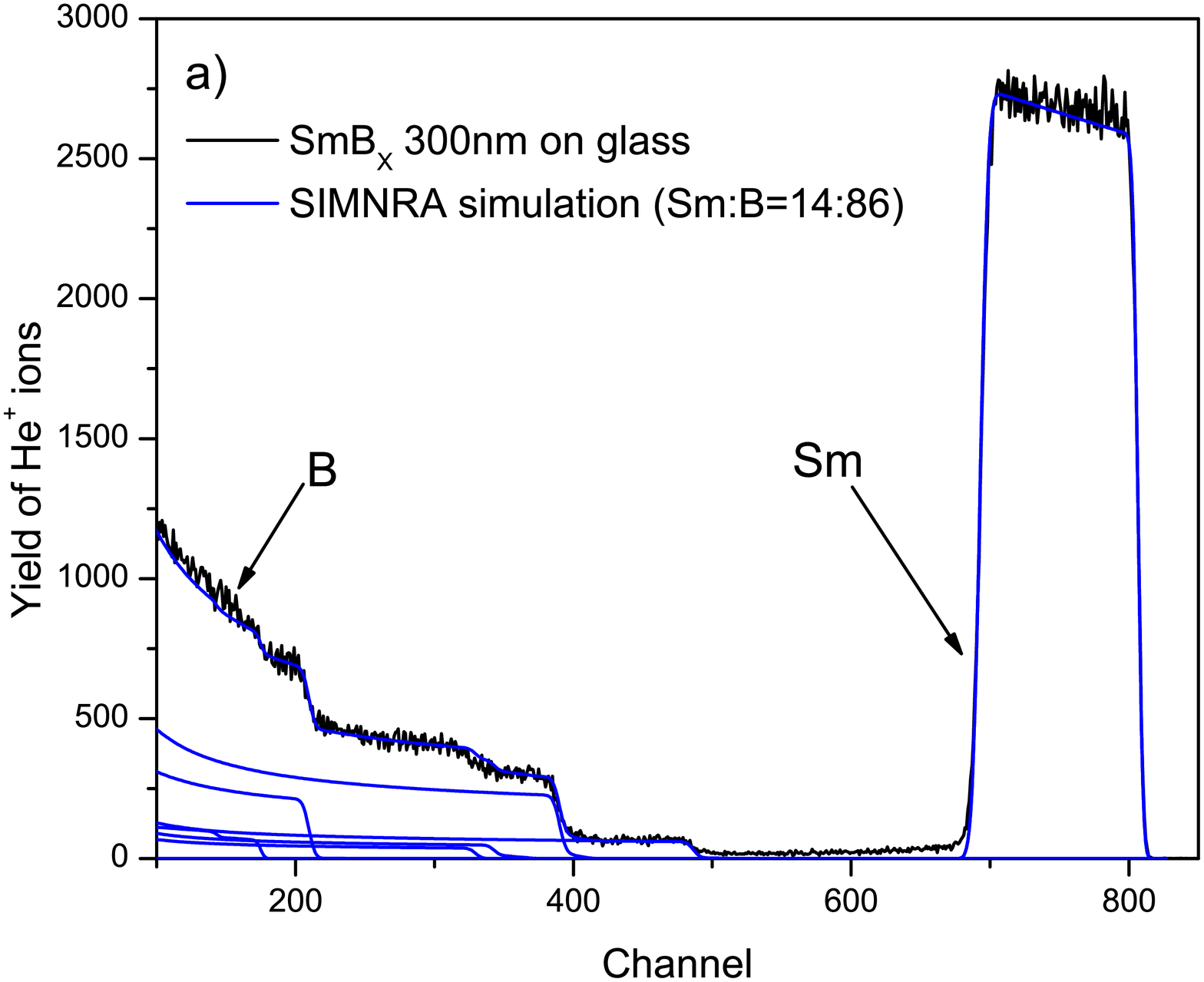}
	\vspace{15pt}
\end{subfigure}
\begin{subfigure}[h]{1\columnwidth}
   \includegraphics[width=0.95\columnwidth]{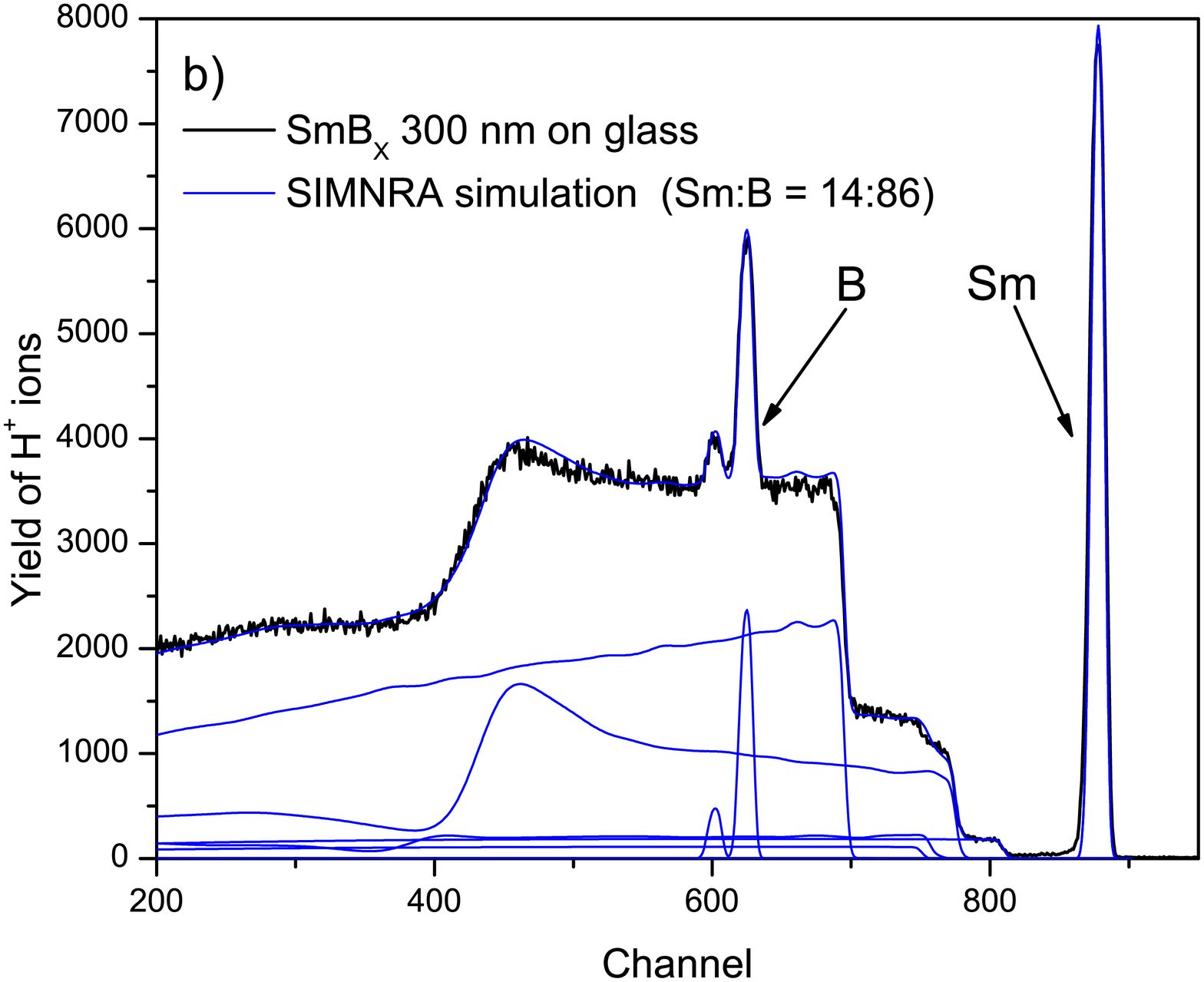}
\end{subfigure}
\caption{RBS spectrum obtained  using He$^+$ (a) and H$^+$ (b) ion beam with energy of 2~MeV.
SIMNRA simulation of RBS spectrum for incident angle of 0$^\circ$ (blue line) are realized by taking into account the same composition of glass substrate.
}
   \label{Fig3} 
\end{figure}

RBS studies were  performed with the aim to determine an exact composition and thickness variation of the prepared thin films. 
			He$^+$ ion beam incoming angle of 0$^\circ$ and 45$^\circ$ were used to increase significantly the depth resolution, thus to see quality of the surface and interface. 
The results for 300~nm film presented in Fig.~\ref{Fig3}a are taken from 0$^\circ$ geometry.
%
The studied films didn't exhibit any differences, when evaluating the spectrum collected under the perpendicular and glancing geometry, or the uncertainties were in the range of RBS precision.
Evaluation of the spectra of the prepared thin film samples using SIMNRA software gave the B:Sm ratio between 85.5:14.5 (5.9) and 86.5:13.5 (6.4), i.e. 86:14 averaged ratio (truncated with regard to method precision). 
	With the aim to increase sensitivity of light element (boron) detection also analogous RBS studies using H$^+$ beam were performed (see Fig.~\ref{Fig3}b). SIMNRA simulation of the spectra (using the same parameters as for He$^+$ ion beam) gave exactly the same (86:14) ratio of B:Sm. 
	The same composition gained from two RBS experiments with different relative sensitivity to boron, represents confirmation that the prepared samples can be considered as  stoichiometric SmB6 thin films (within experimental uncertainty of RBS).

%

		Having found the right 		
		 conditions for the PLD process, 
		 depositions of the SmB$_6$ thin films on several different substrates, 
		namely MgO, sapphire, and (not polished) ultra-low expansion GC were done.
With the aim to obtain crystalline material, now the temperature 
		of the substrates was kept at 600~$^\circ$C which was
		the maximum achievable value in the used PLD apparatus. 
To be able to verify effect of the substrate temperature on the electrical characteristics,
		also the film on the GC substrate kept at ambient temperature was prepared. 
The depositions were done with the wavelength of 532~nm at pulse frequency of 2~Hz, and pulse energy of 59~mJ.
		The deposition rate (determined by a quartz balance) was 0.32~\AA/s.
	The film thickness estimated from parameters of deposition process 
	was approximately 150~nm, 
		except of the films on GC, where lower thickness is expected, because in this case roughly 	the same amount 	of SmB$_6$ material covers larger area due to the non-polished surface.	

\begin{figure}[h]
			\center{
				\resizebox{0.950\columnwidth}{!}{%
  				\includegraphics{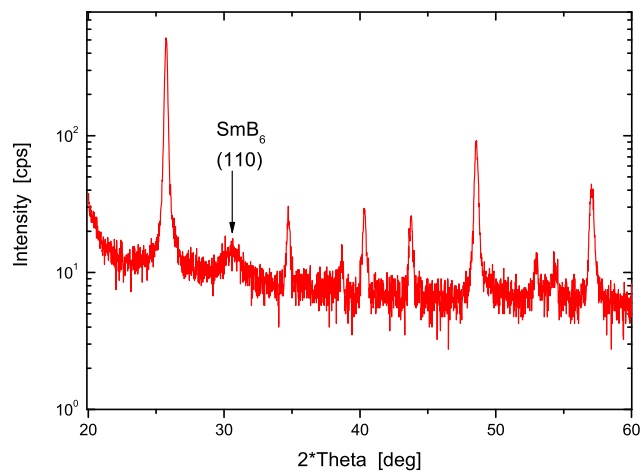}
        }
    	}
	 			\caption{XRD profile of the sample deposited on GC substrate at 600~$^\circ$C.
				The arrow indicates position of (110) peak for SmB$_6$.}
			\label{XRD_1}
		\end{figure}

			XRD studies were performed with the aim to obtain information about structural properties of the prepared films. 
			Measurements of the films deposited on glass substrates at ambient temperature didn't provide  evidence for presence 
			of any texture or crystalline phase.	
			XRD profile of a sample deposited on polycrystalline GC at 600~$^\circ$C is shown in Fig.~\ref{XRD_1}.
			From analysis of this profile it can be deduced that the sample is nanocrystalline with mean grain diameter of about 4~nm and without any distinct texture, or the film can consist of nanocrystalline and amorphous phases. 
			As can be seen in Fig.~\ref{XRD_1}, only one (weak and broad) peak laying at the position of the strongest (110) peak for SmB$_6$  was found in the XRD profile. 
			The other peaks come from the substrate. 
			Results for the sample deposited on GC at room temperature were similar.
		\begin{figure}[h]
			\center{
				\resizebox{0.950\columnwidth}{!}{%
  				\includegraphics{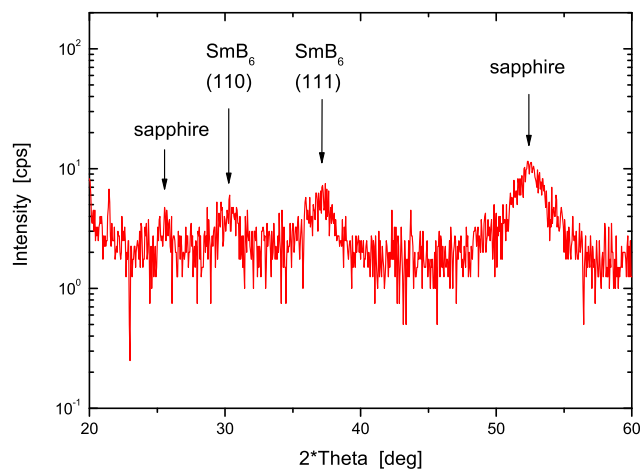}
        }
    	}
	 			\caption{XRD profile taken in BB geometry for the sample deposited on sapphire substrate at 600~$^\circ$C.
				Arrows indicate positions of (110) and (111) peaks for SmB$_6$.}
			\label{XRD_2}
		\end{figure}
		\begin{figure}[h]
			\center{
				\resizebox{0.950\columnwidth}{!}{%
  				\includegraphics{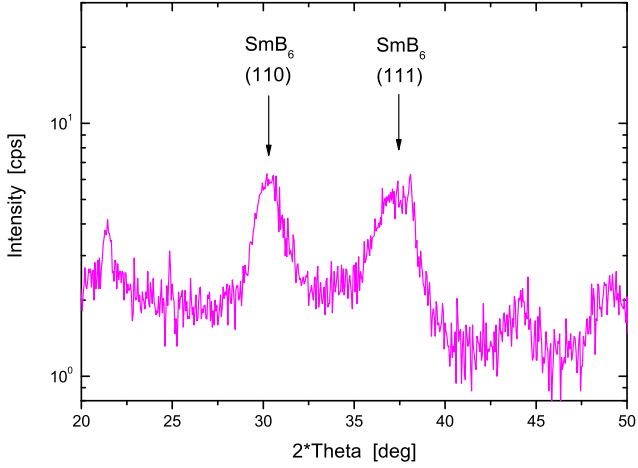}
        }
    	}
	 			\caption{XRD profile taken in GID geometry for the sample deposited on sapphire.
				Arrows indicate positions of (110) and (111) peaks for SmB$_6$.}
			\label{XRD_3}
		\end{figure}

			In accordance with expectations, measurements of the thin films deposited on single crystal substrates were strongly affected by a signal coming from the substrates, therefore we performed measurements of these samples utilizing Bragg-Brentano (BB) geometry and Grazing Incidence Diffraction (GID) geometry. 
 Fig.~\ref{XRD_2} shows XRD profile of a sample deposited on sapphire for BB geometry. 
			The profile reveals a peak which can be assigned to (100) peak for SmB$_6$. 
				The highest intensity peak in the XRD profile is (012) peak at 25,588 degrees, which is very weak such as sapphire substrate shows a miscut of about 1 deg. 
			Two peaks (110) and (111) suggest also polycrystalline films. 
			The XRD profiles shown in Fig.~\ref{XRD_3} were obtained in GID geometry (where an angle of x-ray incidence was equal to 3 deg);
			they confirm above mentioned suggestion, as broad and weak peaks laying at the positions of (110) and (111)   peaks for SmB$_6$ were found there. 
			Unfortunately, we were not able to obtain reliable XRD results for films deposited on MgO as all detected peaks seemed to be connected with the substrate. 	
	

Electrical resistance of the SmB$_6$ target was measured before deposition of the films with the aim to be able 
		to compare it later with resistance of the deposited films. 
The measurement  was performed from room temperature down to 4.2~K,
			while the target was fixed to a sample holder by a Teflon foil,
			and phosphor-bronze springs were used as electrical contacts. 
	(Note that subsequent depositions were done from the side of the target which was not contacted.)
			%
		\begin{figure}[!b]
			\center{
				\resizebox{0.950\columnwidth}{!}{%
  				\includegraphics{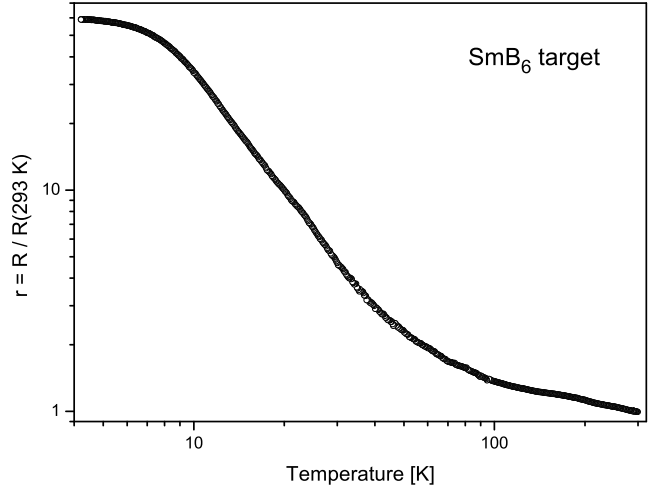}
        }
    	}
	 			\caption{Temperature dependence of the normalized electrical resistance of the SmB$_6$  target.
}
			\label{Fig1}
		\end{figure}
			%
As shown in Fig.~\ref{Fig1}, normalized resistance  $r(T)=R(T)/R($293~K)  of the target reveals typical large 
		resistance increase below 50~K. 
The ratio $R$(5~K$)/R(293$~K)~$\approx 59$ is more than two orders of magnitude lower 
			than for high quality single-crystalline samples  \cite{Batko93, Cooley95,Batkova06}, 
			which has been, however, expected due to larger contribution of surface states 
			in  polycrystalline materials and 
			less purity of the used target material (99.5~\%) 
			in comparison with high-quality single-crystalline samples. 
		%
		
\begin{figure}[!b]
\centering
   \begin{subfigure}[h]{1\columnwidth}
   \includegraphics[width=0.95\columnwidth]{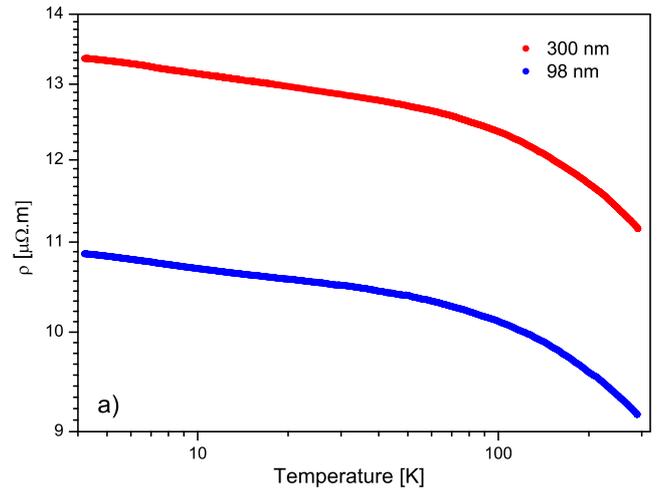}
	\vspace{15pt}
\end{subfigure}
\begin{subfigure}[h]{1\columnwidth}
   \includegraphics[width=0.95\columnwidth]{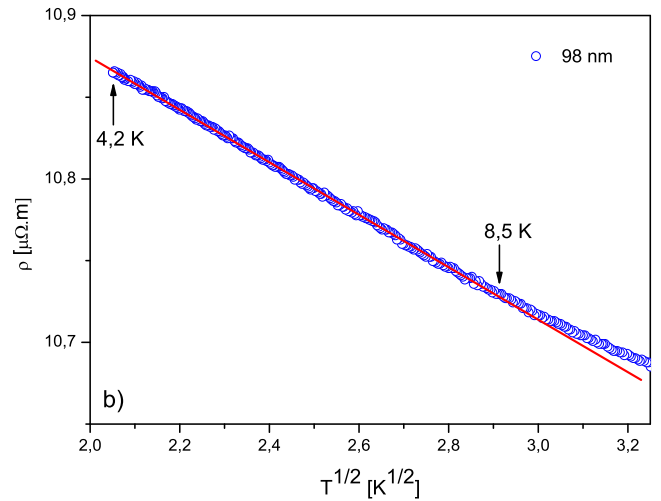}
\end{subfigure}
\caption{Temperature dependencies of electrical resistivity of SmB$_6$ thin films deposited on glass substrates kept at ambient temperature (a) and  $\rho$~{\em versus}~$T^{1/2}$  plot of 98~nm film at lowest temperatures (b).  }
\label{Fig2} 
\end{figure}%

%
As can be seen in Fig.~\ref{Fig2}a, electrical resistivity of the films deposited on glass (at ambient temperature)
 increases with temperature decrease,
			but the largest increase is observed above 50~K, although  dominant resistivity 
			increase for bulk SmB$_6$ samples is expected below 50~K.
The resistivity of the films is more than 3-times higher than the resistivity of bulk SmB$_6$ 
			single crystals \cite{Allen79} or (stoichiometric) thin films reported by other group \cite{Yong14}.
					Such properties can be a consequence of deviation from the stoichiometry, as well as of high structural disorder in the films (e.g. amorphous state). 
	As already mentioned, X-ray scattering studies haven't provided any evidence for presence of texture 
	or crystalline phase(s) in these films, thus supporting the supposition about 
	structural disorder.	
With respect to the theory for disordered metals and heavily-doped semiconductors, 
				which considers effects due to disorder and correlation \cite{Altshuler79}, 
temperature dependence of the resistivity at lowest temperatures can be reasonably expected in the form $\rho(T) = \rho_0 + A T^{1/2}$, where parameter $A$ has negative value, as the contribution to the resistivity due to disorder and correlation effects increases with temperature decrease.
Indeed, the plot of the resistivity data in $\rho$ vs $T^{1/2}$ coordinates,
			shown in Fig.~\ref{Fig2}b reveals 
			that electrical resistance below 9~K  can be described by $T^{1/2}$ law, 
				what we consider as additional indication for presence of structural disorder.

\begin{figure}[h]
			\center{
				\resizebox{0.950\columnwidth}{!}{%
  				\includegraphics{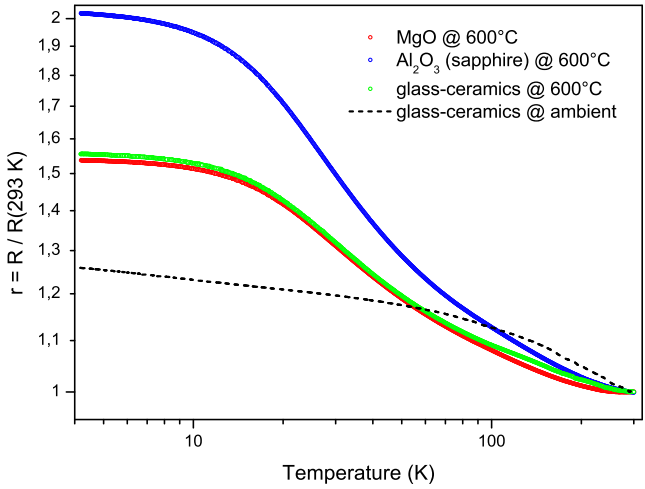}
        }
    	}
	 			\caption{Temperature dependencies of normalized electrical resistances of SmB$_6$ thin films deposited on Sapphire (blue circles), MgO (red circles), and GC (green circles) substrates at substrate temperature of 600~$^\circ$C. The $r(T)$ curve for thin film deposited on the GC substrate at the ambient temperature  is shown for illustration as black dashed line.}
			\label{Fig4}
		\end{figure}

As can be seen in Fig.~\ref{Fig4}, normalized electrical resistance $r(T)=R(T)/R($293~K) of the SmB$_6$  
		thin films deposited 	on substrates kept at 600~$^\circ$C 
		behaves as expected for SmB$_6$, i.e. exhibits significant resistance 
		increase below 50~K, and has the tendency to saturate at lowest temperatures. 
%
On the other hand, the $r(T)$ curve of the thin film deposited on the GC substrate 
		kept at the ambient temperature 
				reveals a different dependence,  which is,  however, 
		qualitatively the same as one for the films deposited  at the ambient temperature on glass 
			substrates, shown in Fig.~\ref{Fig2}. 
%
This confirms that heating the substrate during the deposition is crucial to obtain 
		crystalline SmB$_6$ thin films revealing electrical properties as typical for SmB$_6$,
		while 600~$^\circ$C is sufficiently high temperature.
As can be moreover seen in Fig.~\ref{Fig4}, 
		the normalized resistance is significantly influenced by the substrate material. 
The obtained values of $r($4.2~K) are 2.02, 1.56, and 1.54 for sapphire, GC, and MgO substrate, respectively.
Comparing crystal structures, it seems surprising that the value of $r(4.2$~K) for  MgO  substrate is less
 			than  		for  sapphire, 	because the crystal structure and lattice parameters of MgO are much similar 
				to ones of SmB$_6$ in comparison to sapphire; 
 MgO crystallizes in cubic structure, analogously as SmB$_6$, 
	with the lattice parameter   
		$a_{MgO}$~=~4.216~\AA ~at room temperature, which is very close to that of SmB$_6$, $a_{SmB_6} = $ 4.13~\AA  \cite{Mandrus94},
	while sapphire has hexagonal crystal structure with lattice parameters 
		$a_{sapphire}$~=~4.77~\AA ~
		and $c_{sapphire}$~ =~13.04~\AA. 
It seems that different thermal expansion of individual substrates play more important role in infuencing
	the  properties of  the prepared SmB$_6$ thin films.
Linear expansion coefficient of sapphire (perpendicular to c axis) 
		at room temperature ($6.2 \times 10^{-6}$/K) is less than that of MgO ($8\times 10^{-6}/$K), 
		and is close to one of SmB$_6$, which is approximately $6.5\times 10^{-6}/$K at room temperature, 
		and reaches negative values at low temperatures, at least in the temperature range of 12 - 80~K \cite{Mandrus94}.   
Therefore, properties of SmB$_6$/MgO thin films at room temperatures 
		are expected to be affected by in-plane compressive residual strain
	 induced after the deposition due to contraction of the substrate during cooling from 600~$^\circ$C.  
It is generally known that application of hydrostatic pressure to  SmB$_6$ results in insulator-to-metal transition, 
			which is accompanied  by a strong resistivity decrease  with  increasing pressure 
			at the lowest temperatures \cite{Cooley95,Gabani03}.
		Thus, lower value of $r$(4.2~K) for SmB$_6$/MgO in comparison to
				SmB$_6$/saphire thin films can be explained by greater (positive) linear expansion 
				coefficient of MgO in comparison to that of saphire. 
In addition, consideration that the strain induced in the SmB$_6$ films by MgO substrate
		increases with temperature decrease in a certain region below room temperature,
		allows also to explain a very shallow minimum in the resistivity observed in the vicinity of 270~K.
Opposite effect of substrate on the deposited SmB$_6$ films, 
			namely in-plane tension  
					is expected for  GC substrate having  ultra-low expansion coefficient
			($0.0 \pm 0.2 \times 10^{-7}$/K) at room temperatures,
					and negative one at lower temperatures \cite{Clearceram}, similar to that of SmB$_6$ \cite{Mandrus94}.
The thermal expansion coefficient of sapphire is at the closest to one of SmB$_6$
			and therefore SmB$_6$ thin films on sapphire substrate are supposed to be at least affected by in-plane strains.

\begin{figure}[h]
			\center{
				\resizebox{0.9500\columnwidth}{!}{%
  				\includegraphics{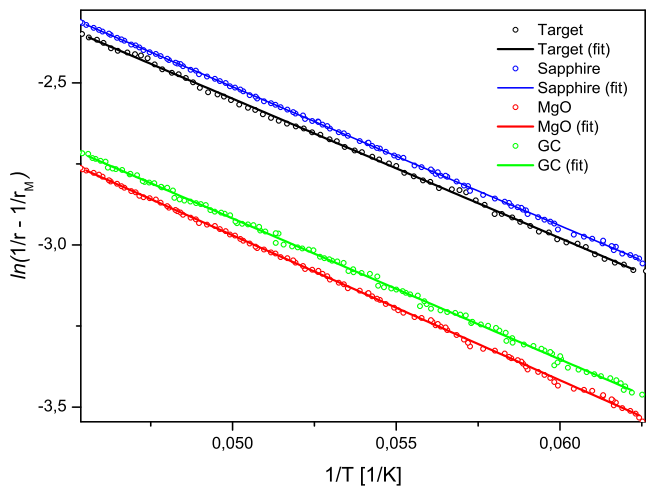}
        }
    	}
	 			\caption{Plots of normalized resistance in coordinates $ln(1/r - 1/r_{M})$ {\em versus} $1/T$ using specific values of fitting parameters $1/r_{M}$ for each sample and the target; values of $1/r_{M}$ are summarized in Table 1. 
}
			\label{Fig5}
		\end{figure}

Let us focus  on electrical properties at the lowest temperatures.
As reported by many authors before, electrical resistivity data of SmB$_6$ at low temperatures can be 
adequately described by the two-channel model of electrical conductivity \cite{Batko93,Cooley95,Syers15,Wolgast13, Petrushevsky17},  
\begin{equation}
	\sigma(T) = \sigma_s + \sigma_b \times \exp(-W_b/k_{B}T),
\label{eq1}
\end{equation}
where $k_B$ is Boltzmann constant, $W_b$ is activation energy of the bulk transport, 
and parameters $\sigma_s$ and  $\sigma_b$ are constants characterizing contributions to the conductivity
from the surface states and the bulk, respectively.  
With the aim to compare electrical properties of the SmB$_6$  target 
			and the thin films deposited from it	we performed  corresponding numerical analysis of the experimental data.
Obtained normalized resistance $r(T)$ of the target and representative thin films plotted in Fig.~\ref{Fig1} and Fig.~\ref{Fig4},
			respectively, were fitted over a temperature interval (the same one for all films and the target), 
			where  Eq.~\ref{eq1} can be reliably applied, and 
			where huge temperature dependence of  the  normalized resistance of the target is observed. 
			%
In accordance with Eq.~\ref{eq1},	the data were fitted by the formula 
\begin{equation}
	1/r(T) = 1/r_M + (1/r_{A}) \times \exp(-W_A/k_{B}T),
\label{eq2}
\end{equation}
where the first term represents temperature non-activated (metallic) contribution,
			and the second one is temperature activated term associated with the bulk.  
Plotting the data 
			in coordinates $ln(1/r - 1/r_M)$ {\em versus} $1/T$ (using specific value of fitting parameter $r_M$ 
			for each individual sample) has provided  straight lines in all cases, 
			at least in the temperature interval of 16~-~22~K, as can be seen in Fig.~\ref{Fig5}. 
Moreover, almost the same slope $W_A/k_{B}$ is observed for all thin films  deposited at 600$^\circ$C, as well as for the target. 
Activation energies $W_A$, determined from the data  
				are summarized in the Table~1 together with other characteristics of the films and the target.
The obtained values of activation energies  are in good agreement with results of other groups \cite{Allen79, Batko93,Cooley95,	Syers15,Petrushevsky17, Wolgast13,Batkova06} 

%
As follows from Tab.~1, the activation energy 
		varies only weakly with the used substrate, and is very close to that of the target.
This is a very important finding especially from technological point of view,   
		because it reveals that bulk properties of the  stoichiometric 
		target material can be effectively transferred by the PLD process to the SmB$_6$ thin films 
		without need for additional boron source. 
Moreover, the bulk properties of the SmB$_6$ thin films do not seem to be 
			essentially affected by the properties of substrates. 
Considering the films deposited on sapphire and MgO 
(of almost the same thickness) one can expect that the bulk contributes  to the total conduction by the same value. 
Thus, we suppose that observed differences in conduction are predominantly caused 
		by different temperature independent contributions ($1/r_{M}$)
				due to different properties of the interface and/or different 
				compressive or tensile strains induced changes 
			in the near surface region, as both, compression \cite{Cooley95, Gabani03}, 
			as well as tensile strain \cite{Stern2017} are parameters strongly affecting electrical properties of SmB$_6$).
In our opinion, the  latter reason is more important, because of the fact that MgO lattice parameters are 
more similar to ones of SmB$_6$, as discussed above.
%
						\setlength{\arrayrulewidth}{0.2mm}
			\setlength{\tabcolsep}{8pt}
			\begin{table}
						\begin{tabular}{ |p{2.2cm}||p{0.9cm}|p{1.2cm}|p{0.7cm}|p{0.7cm}|  }
\hline
  \textbf{Sample}  & \textbf{$r$(4.2~K)}~  & \textbf{$W_{A}$~[meV]}~ & \textbf{1/$r_m$} & \textbf{1/$r_b$ } \\
 \hline
                                   Target      & 59     & 3.70   &   0,023   &  0.670   \\
                                   Sapphire    & 2.02   & 3.69   &   0.504   &  0.690   \\
                                   MgO         & 1.54   & 3.82   &   0.654   &  0.479   \\
 Clearceram\textsuperscript{\textregistered}-Z & 1.56   & 3.75   &   0.647    & 0.447   \\
  \hline
	\end{tabular}
					\caption{Summary of normalized resistances, $r(4.2)$~K, and fitting parameters according to the formula (2) for the polycrystalline SmB$_6$ target and thin films deposited at 600~$^\circ$C on sapphire, MgO and GC.    }
					\label{Tab1}
						\end{table}
%
Nevertheless, additional and more complex studies on wider series 
		of SmB$_6$ thin films are needed; here presented feasibility studies of 
		reproducible grow of stoichiometric SmB$_6$ thin films 
				represent excellent starting information for their realization. 
As we have shown, PLD enables to prepare thin films with the composition and bulk properties 
			that mirror ones of the used target.
Hence, thanks to enabling synthesis of thin films of desired composition from 	
			a single composite or compound target 
				corresponding to the desired composition of the films, 
					the PLD can be very  useful also for 
   studies of doping or deviation from the stoichiometric chemical formula.
		
			%
Also, utilization of PLD technique to deposit SmB$_{6}$ (or SmB$_{6}$-based) electrodes could enable (in combination with another deposition techniques, e.g. magnetron sputtering)  fabrication of tunnel structures and
consequent realisation of complex (planar) tunnelling spectroscopy studies of effect of composition on electronic properties of surface states in SmB$_{6}$-based materials. 
In fact, mentioned approach could help to eliminate some technology problems 
arising in experiments
if planar tunnel junctions are fabricated on the surface of SmB$_6$ single crystals \cite{Sun17}, 
and it could help to extend such studies also for composition similar (SmB$_{6}$-based) materials, 
not (easily) achievable in a bulk-crystal form.  
%

\section{Conclusions}
SmB$_6$  thin films were successfully prepared by PLD 
			from a single stoichiometric polycrystalline SmB$_6$ target. 
	It was demonstrated that  using the right conditions of PLD, escape of boron can be avoided 
			and there is no need for additional boron target.
 Performed RBS studies have shown that PLD process enables a growth of SmB$_6$ thin films
			without any measurable deviation from the composition of the target. 
		XRD studies provided indication that the prepared films are nano- or polycrystalline. 
Bulk electrical properties of the films very close to those of the target 
			were reached by deposition on  MgO, sapphire, and glass-ceramics substrates 
			heated up to 600~$^\circ$C.
The highest ratio $R(4.2)/R(293)$ was observed for SmB$_6$ films on sapphire, 
		what we predominantly associate with the fact that thermal expansion parameter of sapphire 
		is the 	closest to that of SmB$_6$ in comparison to other used substrates. 
This might indicate preference of sapphire for potential applications. 
Also cost effective glass-ceramics substrates seem to be  very promising alternative,  
			because SmB$_6$ films deposited on this substrate exhibit
			a higher resistance increase upon cooling in comparison with the films deposited 
			on (frequently used) single crystalline MgO substrate.

\section{Acknowledgments}
This work was supported by the Slovak Scientific Agency VEGA (Grant No.~2/0015/17), by Czech Science Foundation (GACR No. P108/12/G108) and by European Regional Development Fund (ITMS 26110230097). 
The RBS studies were carried out at the CANAM (Centre of Accelerators and Nuclear Analytical Methods) 
infrastructure LM 2015056. M.B and I.B. thank L. Havela of Charles University in Prague for his helpful suggestions and valuable comments.

%
%

\section*{References}










\end{document}